\documentclass[twocolumn,           
               showpacs,            
               preprintnumbers,     
               aps,                 
               prd,                 
               a4paper,             
               groupedaddress,  
               nofootinbib,         
               tightenlines,        
               floats,floatfix      
               ]{revtex4}

\usepackage{graphicx}
\usepackage{subfigure}

\usepackage{latexsym}
\usepackage{amsmath,amssymb}        
\usepackage[draft=false]{hyperref}
\usepackage{threeparttable}

\newcommand{\abs}[1]{\left\vert#1\right\vert}

\input colordvi

\begin{document}



\title{Cosmology at the boundary of de Sitter using the dS/QFT correspondence}
\affiliation{Astronomy Centre, University of Sussex, Brighton BN1 9QH,
  United Kingdom}
\author{Mafalda Dias}
\affiliation{Astronomy Centre, University of Sussex, Brighton BN1 9QH,
  United Kingdom}
\date{\today}


\begin{abstract}

Using the dS/QFT correspondence in the context of inflation allows for the study of interesting, otherwise inaccessible physics. In particular, by studying inflation \textit{via} its dual field theory at the boundary of the de Sitter space, it may be possible to study a regime of strongly coupled gravity at early times. The purpose of this work is to completely express cosmological observables in terms of the free parameters of a dual field theory and to compare them with CMB data. In this way, constraints on the observational parameters constrains the validity of the strongly coupled inflation picture by imposing limits on the parameters of the field theory.
The fit with data defines a limit for the consistency and validity of the approach taken and shows that, within this limit, the model is almost unconstrained, but quite predictive, producing power spectra of density perturbations extremely near scale invariance.

\end{abstract}

\maketitle


\section{Introduction}

One of the most interesting ideas that emerged from string theory is the AdS/CFT correspondence \cite{david41, david42, novo, david43}. This states that a theory with gravity in a $D$-dimensional anti de Sitter space is dual to, and can be described by, a conformal field theory 
that lives in the ($D$-1)-dimensional boundary of this space. Furthermore, this holographic duality is weak/strong, meaning that when the theory with gravity is in a strongly coupled regime, its dual field theory will be in a weakly coupled one, and vice-versa. This is an extremely curious and powerful correspondence. The idea initially came to light from the study of the large $N$ limit of $SU(N$) theories. It turns out that in this limit, $SU(N$) theories behave like string theories one dimension higher, with a coupling constant of 1/$N$ (see for example Ref.~\cite{coleman}).  String theory contains gravity, so the large $N$ limit connects $SU(N$) field theories with gravity. A concrete realisation of this connection was made explicit by Maldacena for the case of the perturbative type IIB string theory in a AdS$_{5} \times \rm{S}_{5}$ space with $N$ branes. Such theory, in the near boundary limit and for large $N$, is dual to a $\mathcal{N}=4$ $SU(N$) super-Yang-Mills field theory that lives in the boundary of the AdS space \cite{david41, david42, novo, david43}.

In fact, this idea of holography is known outside the scope of string theories. The study of black hole thermodynamics shows that the entropy, and so the degrees of freedom, of a theory with gravity acting on a space $\Gamma$ needs to be proportional to the area of the boundary of $\Gamma$. In the case of field theories, though, the degrees of freedom are proportional to the volume of the space on which they act. This can be understood in a holographic way: a theory with gravity has its degrees of freedom on the boundary of $\Gamma$ and so there will be a dual field theory that acts on this boundary that has the same degrees of freedom. The dual field theory of the boundary can then describe all the physics of $\Gamma$ (for detailed discussion, see Ref.~\cite{susskind}). That this idea, based on simple concepts like thermodynamics, fits so perfectly well with the duality realised in string theory shows an amazing consistency of such principles. 

Although the exact correspondence is only known for type IIB string theories in 5 dimensional AdS space, one can expect that this correspondence exists for any theory of gravity in AdS space \cite{david45}. In fact, one can construct a dictionary that relates correlation functions of theories in the $D$-dimensional AdS bulk with their dual correlators in the boundary field theory \cite{david42, novo}. Furthermore, given that AdS spaces are related to dS spaces by a double Wick rotation, one can hope to be able to extend this dictionary to dS/QFT correspondences \cite{david2, novo, david, david36, david37, david38}. In this form, holographic principles can be applied to cosmology, since inflation behaves like a quasi de Sitter space. 

In this paper, our intention is to study inflation in the light of these duality ideas. It seems almost irresistible to undertake this task. If there is a field theory dual to cosmological inflation, and if there is a dictionary that relates correlation functions of both, it is in principle possible to evaluate correlators associated with cosmological observables in the bulk and boundary. The comparison of these with observations can be a powerful consistency check of the dS/QFT correspondence.

Beside the conceptual interest that a holographic study of inflation can bring, the duality also opens the door to useful computational techniques. In particular, some correlation functions like the bi-spectrum or tri-spectrum of the density perturbations can be easier to evaluate in the boundary dual field theory than in the full bulk cosmology. The evaluation of such parameters is increasingly necessary with the advent of new data from missions like Planck, and there has been a lot of work done in this direction \cite{david2, novo, david, holo3}.

On the other hand, as mentioned, the duality is weak/strong. In practical terms, for particle physics, this is one of the most useful characteristics of holography, since it provides an approach to otherwise analytically inaccessible physics. A theory in a regime where perturbative analysis is not possible is unveiled \textit{via} its dual theory, which is itself in a weakly coupled regime and can then be treated analytically.  

In this paper, we want to make use of this feature of holography for cosmology, by computing all observables on the boundary of the inflationary de Sitter space, in the same way as Ref.~\cite{holo, holo2, holo3}. Assuming that the boundary field theory can be treated perturbatively, we impose that the bulk theory during inflation cannot be described by a perturbed FRW background since it is in a regime where gravity is strongly coupled. This sort of physics was impossible to constrain without using dS/QFT techniques, but by comparing observational parameters with data, we can define limits to the viability of this scenario. This is the main goal of the current work.

\section{dS/QFT correspondence and Cosmology}

\subsection{dS/QFT dictionary}

The idea that a correspondence between a theory of gravity in a de Sitter space and a field theory at its boundary should exist is an extension of the known correspondence in anti de Sitter spaces. The duality in the case of an anti de Sitter space is only realised completely for the very particular case of type IIB string theory in a AdS$_{5} \times S_{5}$ background \cite{david41}, however it is possible that this correspondence may hold more generally \cite{david45}. The extrapolation to de Sitter spaces is founded on the fact that de Sitter and anti de Sitter spaces are related by a double Wick rotation of the radial and time coordinates, however there is no guarantee that such a rotation holds for the boundary theory as well \cite{sus}. The correspondence that will be used is weaker and was presented for the first time in Ref.~\cite{david2}. What is stated is that there is a correspondence between bulk and boundary that is expressed via correlation functions of a theory of gravity of the bulk and a field theory at the boundary (this is used in, for example, Ref.~\cite{david, david36, david37, david38}). This is equivalent to the approach of Ref.~\cite{holo, holo2, holo3}, where the Wick rotation is interpreted as an analytical continuation between an anti de Sitter space and its de Sitter domain wall. In this case, all computations regarding the correspondence are done between AdS/QFT and the results are brought to the domain wall case by applying this continuation. 

To briefly understand how the correspondence works, let us consider a de Sitter space with a single massless scalar field. 
The de Sitter metric can be written as:
\begin{equation}
ds^{2} = a^{2}(\eta) [ -d\eta^{2} + \delta_{ij}dx^{i}dx^{j}]
\label{metric}
\end{equation}
where $\eta$ is the conformal time defined as $\eta = \int dt/ a(t) $ and the scale factor behaves as $a = e^{Ht}=-(H\eta)^{-1}$. The far past corresponds to $t \rightarrow  -\infty$ and $\eta \rightarrow -\infty$, whereas the far future corresponds to $t \rightarrow +\infty$ and $\eta \rightarrow 0^{-}$. The boundary of the de Sitter space, $\partial$dS, where we expect the dual field theory to live, coincides with this far future. The presence of the scalar field, which respects the field equation
\begin{equation}
\ddot{\phi} + 3H\dot{\phi} = -\frac{dV}{d \phi}.
\end{equation}
breaks the de Sitter symmetries, but at $\partial$dS the metric is asymptotically de Sitter and $V$ is negligible. 
So, near the boundary, the solution to this equation has the form (as can be seen by looking at powers of $\eta$)
\begin{equation}
\phi \sim \hat{\phi}e^{0Ht} + \bar{\phi}e^{-3Ht}.
\label{solution}
\end{equation}
Asymptotically, in $\partial$dS, the solution is just $\phi \sim \hat{\phi}$. How can we relate this bulk theory to a field theory in $\partial$dS? The dS/CFT correspondence \cite{david2, david32} states that the wave function of the bulk gravity theory as it approaches the boundary is given by the partition function of the boundary dual field theory:

\begin{align}
\Psi_{dS}=Z_{QFT}[\phi_{\partial \rm{dS}}]
\end{align}
The wave function, provided that the bulk curvature is sufficiently small,  can be evaluated from the action of the classical solution at the boundary. Such an action will then be a functional of $\hat{\phi}$ which implies that the partition function of the field theory should be a functional of the same $\hat{\phi}$. The partition function is a generating function of the field theory, \textit{i.e.}, a functional of sources from which correlators can be computed. So, one can identify $\hat{\phi}$ as a source for an operator $\mathcal{O}$ of the dual field theory \cite{david42}. Then this operator $\mathcal{O}$ is dual to the scalar field $\phi$.

\begin{align}
Z_{QFT}[\hat{\phi}]= \Big{\langle} \exp \left( \int_ {\delta dS} d^{3}x \  {\mathcal O}\hat{\phi} \right) \Big{\rangle}_{QFT} \simeq e^{iS[\hat{\phi}]}
\end{align}

Correlation functions of $\mathcal{O}$ can be obtained by taking the derivative of the partition function in terms of the source and setting it to zero, as usual.

\begin{align}
\langle {\mathcal O}(x_{1}) ... {\mathcal O}(x_{n})\rangle = \left. \frac{\delta^{n}Z[\hat{\phi}]}{\delta \hat{\phi}(x_{1}) ...\delta\hat{\phi}(x_{n})}\right|_{\hat{\phi}=0}
\end{align}

These two expressions define a dictionary between correlators of $\phi$ from the bulk and $\mathcal{O}$ from the boundary. As an example, the 2-point function for the single scalar field is \cite{david2, david38}
\begin{equation}
\langle {\phi(k_{1}) \phi(k_{2})\rangle = -\frac{1}{2 \rm{Re} \langle {\mathcal O}(k_{1}) {\mathcal O}(k_{2})\rangle}}
\end{equation}

Going back to expression (\ref{solution}), it makes sense that $\hat{\phi}$ should be related to deformations of the boundary field theory \textit{via} the operator $\hat{\phi} \mathcal{O}$. In fact, the solution $\phi \sim \hat{\phi}$, as one approaches $\partial$dS, doesn't decay, which means that it corresponds to an infinite energy excitation of the bulk wave function. This can be seen as a deformation of the gravitational background, in accordance with its interpretation for the dual field theory. In the same way, one can identify the meaning of $\bar{\phi}$. It is associated with a solution that decays as one approaches $\partial$dS, \textit{i.e.}, a normalisable solution and so finite energy excitation of the bulk gravity. It can then be connected to a finite energy excitation of the dual operator $\mathcal{O}$ on the boundary theory, in other words its VEV. In fact, it can be shown that $\bar{\phi}$ is directly related to $\langle{\mathcal{O}\rangle}$ \cite{david42}; it is generally referred as the \textit{response}.

\subsection{Cosmological observables}

In this work, we are interested in studying inflation using the dS/QFT correspondence by computing cosmological observables and comparing them with observations. In the case of the present work, by cosmological observables we mean the power spectrum of scalar and tensor perturbations from which we can compute parameters directly comparable with observations, like the spectral index $n_{s}$, ratio of tensor to scalar fluctuations $r$, and the running of the spectral index. 
In this subsection we present the quantities we aim to compute with the holographic dictionary. We follow the notation from Ref.~\cite{holo}.

Linearly perturbing the metric (\ref{metric}), we can see that the fluctuations can be defined by two functions, $\zeta$ and $\gamma_{ij}$. $\zeta$ represents curvature perturbations on comoving hypersurfaces whereas $\gamma_{ij}$ is associated to tensor modes of the fluctuations. Since the correlation functions are going to be evaluated far outside the horizon (at the future boundary) one can expect them, even if they are not constant after horizon crossing,  to approach a constant late-time solution. In the light of this argument, the power spectra can be written as:
\begin{equation}
\Delta_{S}^{2} = \frac{k^{3}}{2\pi^{2}}\langle{ \zeta(k) \zeta(-k)\rangle} = \frac{k^{3}}{2\pi^{2}} |\zeta_{k(0)}|^{2}
\end{equation}
\begin{equation}
\Delta_{T}^{2} = \frac{k^{3}}{2\pi^{2}}\langle{ \gamma_{ij}(k) \gamma_{ij}(-k)\rangle} = \frac{2 k^{3}}{\pi^{2}} |\gamma_{k(0)}|^{2}
\end{equation}
where $\zeta_{k(0)}$ and $\gamma_{k(0)}$ are the constant late-time values of the functions, and $k$ the wavenumber of the perturbations. Applying the canonical commutation relations, it is possible to normalise the mode functions, obtaining this way the Wronskian conditions
\begin{align}
&i= \zeta_{k}\Pi^{\zeta *}_{k} - \Pi^{\zeta}_{k}\zeta^{*}_{k}\\
&i/2= \gamma_{k}\Pi^{\gamma *}_{k} - \Pi^{\gamma}_{k}\gamma^{*}_{k}
\end{align}
where $\Pi_{k}^{\zeta} = -2M_{\rm{Pl}}^{2}a^{3}\dot{\zeta}_{k}\dot{H}/H^{2}$ and $\Pi_{k}^{\gamma}=(1/4)M_{\rm{Pl}}^{2}a^{3}\dot{\gamma}_{k}$, with $M_{{\rm Pl}}^{-2}=8\pi G$ and $\hbar $ set to unity.

We can rewrite the power spectra in a more convenient way for the computations ahead by defining the linear response functions $E$ and $\Omega$:
\begin{align}
\Pi_{k}^{\zeta} = \Omega \zeta_{k} \quad \quad  \Pi_{k}^{\gamma}= E \gamma_{k}.
\end{align}
Inserting these definitions in the Wronskian conditions, we have the power spectra
\begin{align}
\Delta_{S}^{2} = \frac{-k^{3}}{4\pi^{2}  {\rm Im}   \Omega_{k(0)}}\\
\Delta_{T}^{2} = \frac{- k^{3}}{2\pi^{2}  {\rm Im}  E_{k(0)}}
\end{align}

\subsection{dS/QFT dictionary for cosmological observables}

Since the aim is to compute the power spectra of metric perturbations, it makes sense that instead of working with the 2-point function of the inflaton field, we should work straightaway with the correlation functions of these perturbations. Just as in subsection II.A, in this subsection a dictionary linking perturbations with their dual operator is presented. The procedure is identical, starting with the study of the asymptotic behaviour of the metric near $\partial$dS \cite{holo}:
\begin{equation}
ds^{2} = a^{2}(\eta) [ -d\eta^{2} + g_{ij}dx^{i}dx^{j}]
\end{equation}
where
\begin{equation}
g_{ij} \sim \hat{g}_{ij} +\bar{g}_{ij}e^{-3Ht}.
\end{equation}
In this case, for the same reasons as in the scalar field analysis,  $\hat{g}_{ij}$ and $\bar{g}_{ij}$ can be identified as a source and a response, respectively. But in this case, what is $\mathcal{O}$ the dual operator associated with the perturbations in the metric? It can be shown \cite{Tij} that in fact:
\begin{equation}
\langle{ T_{ij}\rangle} = -\frac{3M_{{\rm Pl}}^{2}}{2}\bar{g}_{ij}
\label{tij}
\end{equation}
where $T_{ij}$ is the stress-energy tensor of the dual QFT at the boundary $\partial$dS. The response $\bar{g}_{ij}$, as expected, is directly related to the 1-point function of the dual operator $T_{ij}$; the power spectra will then be related to the 2-point correlation function of this $T_{ij}$. 

In the same way that perturbations in the metric and the inflaton field are related, so $T_{ij}$ and $\mathcal{O}$ need to be related by the Einstein equations. 
Perturbing $T_{ij}$ to first order in the operator $\mathcal{O}$ gives:
\begin{align}
\label{deltatij}
\delta \langle{T_{ij}\rangle} =& - \int d^{3}y \sqrt{\hat{g}} \left( \frac{1}{2} \langle{ T_{ij}(x) T_{kl}(y)\rangle} \delta \hat{g}^{kl} + \right.\\
&\left. \langle{ T_{ij}(x) \mathcal{O} (y)\rangle} \delta \hat{\phi}(y) \right). \nonumber
\end{align}
It is helpful to re-express $\langle{ T_{ij}(k) T_{kl}(-k)\rangle} $ in terms of projection operators, in this way:
\begin{equation}
\langle{ T_{ij}(k) T_{kl}(-k)\rangle} = A(k) \Pi_{ijkl} + B(k) \pi_{ij}\pi_{kl}
\end{equation}
where
\begin{align}
&\Pi_{ijkl} = \frac{1}{2}(\pi_{ik}\pi_{lj} + \pi_{il}\pi_{kj} - \pi_{ij}\pi_{kl}) \\
&\pi_{ij}= \delta_{ij} - \frac{k_{i}k_{j}}{k^{2}} \nonumber
\end{align}
Perturbing explicitly the left-hand side of expression (\ref{deltatij}) using (\ref{tij}) and matching the result with the right-hand side, we get the relations (using the definitions of the linear response functions $E$ and $\Omega$) \cite{holo}:
\begin{equation}
A(-ik)= 4 E_{k(0)} \quad \quad B(-ik)=\frac{1}{4} \Omega_{k(0)}
\end{equation}
which leads to the result,
\begin{align}\label{resultado}
\Delta_{S}^{2}(k) = \frac{-k^{3}}{16\pi^{2} {\rm Im} B(-ik)}\\  \nonumber
\Delta_{T}^{2}(k) = \frac{-2k^{3}}{\pi^{2} {\rm Im} A(-ik)}
\end{align}
with which we can match the power spectra of the metric perturbations of the bulk with the 2-point function of the stress-energy tensor of the boundary quantum field theory.

\section{Cosmology at the dual boundary}

As mentioned, a lot of work has been done applying dS/QFT to cosmology, but since we do not know what the boundary field theory looks like, the general approach is to do all computations based on the bulk theory \cite{david2, david, david36, david37, david38}. The bulk theory in this case is just standard inflation, and all holographic analysis of the dual theory needs to be done in such a way that the standard known results of the bulk are recovered. This approach gives consistency checks to the duality dictionary and can be helpful for some computations, but it doesn't unveil any new physics or bring any new real knowledge on the nature of inflation. 

However, these gravity/field theory dualities are often used in particle physics because they provide a tool for calculations in certain physical regimes otherwise unaccessible analytically, because of their weak/strong nature. This just means that when the bulk theory is in a strongly coupled regime, the dual boundary theory is in a weakly coupled one and vice-versa. If one tries to analyze cosmology from its dual boundary theory using perturbative analysis, or in other words, assuming that it is weakly coupled, one is then studying correspondingly a regime where bulk gravity is strongly coupled. This would mean that at a very early stage, the universe couldn't be described by the usual geometric picture of fluctuations in the background metric, but information from it can still be computed \textit{via} the dynamics of its dual field theory. This is the proposal of Ref.~\cite{holo, holo2, holo3} and is the approach that is intended to be taken in this work. By comparing the power spectra computed from the dual QFT with observational data, we wish to put some constraints on this scenario.

Clearly, such a non-geometric phase had to end smoothly, giving rise to the standard Big Bang picture, where the description of a FRW geometry with inhomogeneities holds. These perturbations of the metric can be observed and measured. One important question that such an approach needs to consider is then whether or not this transition is possible and reasonable. The end of the strongly coupled regime of the bulk cosmology is equivalent to the end of inflation, so one can see this problem as an analogous to the reheating in the standard scenario. It is not yet known how to make this connection between the non-geometric phase to the subsequent hot Big Bang evolution. In this paper, it is assumed that $\zeta$ and $\gamma$ are conserved through the transition, so that their correlations are set during the non-geometric era and can be computed using the boundary theory.
 \\

A clear challenge for the study of cosmology from the boundary is, as said, that the dual field theory is not known. In fact, it might not even exist since the Wick rotation between dS and AdS spaces might not hold for the boundaries \cite{sus}; the duality for the de Sitter case relates correlation functions \textit{via} the dS/QFT dictionary. A way to go around this problem is to consider that the rotation that relates dS with AdS is an analytical continuation of some variables and that it can be applied as well to correlation functions of the dual boundary theories \cite{holo, holo2, holo3}. In this way, assuming that there is a QFT dual to the AdS space corresponding by analytical continuation to the dS cosmology, one can make all computations with this theory. The correlation functions dual to the cosmology are then obtained by applying the continuation to the functions from this QFT. 

The correct analytic continuation relating dS with AdS can be expressed through the change of variables \cite{holo}:
\begin{align}
N^{2}=-\bar{N}^{2} \quad \quad k= -i\bar{k}
\end{align}
where barred quantities are AdS dual, and unbarred are dS dual; $N$ is the rank of the field theory. But the question remains: what is this QFT? If we take an \textit{ad-hoc} very general theory, allowing for a wide range of fields, we can hope that comparison with observations can impose serious constraints and give an insight on how the real QFT looks. So, by constraining the field content and coupling constant we hope to have constraints on the scenario of strongly coupled gravity at early times. The chosen \textit{ad-hoc} theory is a 3-dimensional $SU(\bar{N}$) Yang-Mills theory, coupled to scalars and fermions all transforming in the adjoint representation. Yang-Mills theories are the prototype theories arising from AdS/CFT computations. The action of such theory looks like:
\begin{align}\label{action}
&S=\frac{1}{g^{2}_{{\rm YM}}}\int d^{3}x \ \rm{tr}\left[ \frac{1}{2}F^{I}_{ij}F^{Iij} + \frac{1}{2}(D\phi^{J})^{2} + \frac{1}{2}(D\chi^{K})^{2} \right.\\
& + \bar{\psi}^{L}\displaystyle{\not}{D} \psi^{L} + \lambda_{M_{1}M_{2}M_{3}M_{4}}\Phi^{M_{1}}\Phi^{M_{2}}\Phi^{M_{3}}\Phi^{M_{4}} +\nonumber \\
&\left. \mu^{\alpha \beta}_{ML_{1}L_{2}}\Phi^{M}\psi_{\alpha}^{L_{1}}\psi_{\beta}^{L_{2}}\right] \nonumber
\end{align}
where there are $\mathcal{N}_{A}$ gauge fields $A^{I} (I=1, ..., \mathcal{N}_{A})$ associated with the field strength $F_{ij}$, $\mathcal{N}_{\phi}$ minimal scalars $\phi^{J} (J=1, ..., \mathcal{N}_{\phi})$, $\mathcal{N}_{\chi}$ conformal scalars $\chi^{K} (K=1, ..., \mathcal{N}_{\chi})$ and $\mathcal{N}_{\psi}$ fermions $\psi^{L} (L=1, ..., \mathcal{N}_{\psi})$. $\Phi^{M}$ represents the interaction term between minimal and conformal scalars $\Phi^{M}= (\{ \phi^{J}\}, \{\chi^{K}\})$. In three dimensions, the coupling   $g_{{\rm YM}}^{2}$has dimensions of energy.
The stress-energy tensor looks like \cite{holo}:
\begin{align}
&T_{ij}=\frac{1}{g^{2}_{{\rm YM}}} \rm{tr}\left[ 2F^{I}_{ik}F^{Ik}_{j} + D_{i}\phi^{J}D_{j}\phi^{J} + D_{i}\chi^{K}D_{j}\chi^{K} \right.\\
& - \frac{1}{8}D_{i}D_{j}(\chi^{K})^{2} + \frac{1}{2}\bar{\psi}^{L}\gamma_{(i}\stackrel{\leftrightarrow}{{D}}_{j)} \psi^{L}  \nonumber \\
& -\delta_{ij}\left(  \frac{1}{2}F^{I}_{kl}F^{Ikl} + \frac{1}{2}(D\phi^{J})^{2} + \frac{1}{2}(D\chi^{K})^{2} - \frac{1}{8}D^{2}(\chi^{K})^{2}\right. \nonumber \\
& \left. \left.+ \lambda_{M_{1}M_{2}M_{3}M_{4}}\Phi^{M_{1}}\Phi^{M_{2}}\Phi^{M_{3}}\Phi^{M_{4}} + \mu^{\alpha \beta}_{ML_{1}L_{2}}\Phi^{M}\psi_{\alpha}^{L_{1}}\psi_{\beta}^{L_{2}}\right)\right] \nonumber
\end{align}

Now it is only required to compute the 2-point function of this $T_{ij}$. Working perturbatively, the leading contribution to this comes from 1-loop diagrams. In this case, it is necessary to sum over contributions of all fields and permutations, each diagram having a contribution of order $\sim \bar{N}^{2} \bar{k}^{3}$. This yields to the result \cite{holo}:
\begin{align}
&A(\bar{k})= C_{A}\bar{N}^{2}\bar{k}^{3} \\
&B(\bar{k})= C_{B}\bar{N}^{2}\bar{k}^{3} \nonumber
\end{align}
where
\begin{align}
&C_{A}= (\mathcal{N}_{A}+\mathcal{N}_{\phi}+\mathcal{N}_{\chi}+2\mathcal{N}_{\psi})/256 \\
&C_{B}= (\mathcal{N}_{A}+\mathcal{N}_{\phi})/256 \nonumber
\end{align}
The first approximation to the power spectra is obtained by substituting these expressions in (\ref{resultado}). In terms of unbarred variables:
\begin{align}
&\Delta_{S}^{2}(k) = \frac{1}{16 \pi^{2} N^{2} C_{B}} + O(g_{{\rm YM}}^{2}(k^{*})/k)\\
&\Delta_{T}^{2}(k) = \frac{2}{ \pi^{2} N^{2} C_{A}}+ O(g_{{\rm YM}}^{2}(k^{*})/k). \nonumber
\end{align}
Deviations from scale invariance arise from the correction of 2-loop diagrams.
After renormalization at scale $\bar{k}^{*}$, these 2-loop diagrams contribute with a factor $\sim \bar{N}^{3}g_{{\rm YM}}^{2}(k^{*}) \bar{k}^{2} \ln(\bar{k}/\bar{k}^{*})$ \cite{holo}. Collecting 1- and 2-loop contributions, the power spectra (expressed in unbarred variables) look like \cite{holo}:
\begin{align}
\label{powers}
\Delta_{S}^{2}(k) = &\frac{1}{16 \pi^{2} N^{2} C_{B}}\left[1 - D_{B}g_{{\rm YM}}^{2}(k^{*})\frac{N}{k}\ln\frac{k}{k^{*}} \right. \nonumber \\
& \left. + O(g_{{\rm YM}}^{4}(k^{*})N^{2}/k^{2})\right]\\
\Delta_{T}^{2}(k) = &\frac{2}{ \pi^{2} N^{2} C_{A}}\left[1 - D_{A}g_{{\rm YM}}^{2}(k^{*})\frac{N}{k}\ln\frac{k}{k^{*}} \right. \nonumber \\
& \left. + O(g_{{\rm YM}}^{4}(k^{*})N^{2}/k^{2})\right] \nonumber
\end{align}
where $g_{{\rm YM}}(k^{*})$ refers to the coupling evaluated at the renormalisation point, and $D_{A}$ and $D_{B}$ are constants that depend on the field content, analogous to $C_{A}$ and $C_{B}$. In this case, $|D_{A}|$ and $|D_{B}|$ are naturally of order unity \cite{holo}. A better way to understand the scale $k^{*}$ in this scenario is by reading it as the scale at which infra-red effects become important for the power spectrum. 

We can identify the dimensionless effective coupling constant of the perturbative expansion: $g_{{\rm eff}}^{2}=g_{{\rm YM}}^{2}(k^{*})N/k$. As $k$ gets smaller, $g_{\rm{eff}}^{2}$ increases until it gets of order $ \sim 1/\ln(k/k^{*})$ at which point the 2-loop corrections become important and scale invariance is broken. 

\section{Comparison with observations}

The main objective of this work is to constrain the scenario developed in section III. This is achieved by constraining the free parameters in the expressions for the scalar and tensor power spectra of perturbations (\ref{powers}), which are the field content, encoded in the constants $C_{A}, C_{B}, D_{A}$ and $D_{B}$ and the choice of scale $k^{*}$ together with the value of the effective coupling constant $g_{{\rm eff}}$ related to this $k^{*}$. 
We use CosmoMC \cite{cosmomc} to fit these expressions with WMAP 7 year release data \cite{wmap}. For this purpose, it is useful to re-express them as:
\begin{align} \label{lalala}
&\Delta_{S}^{2}(k) = \Delta_{S}^{2}(k_{0})\left[1 + S\frac{k_{0}}{k} e^{\alpha}\left(\ln\frac{k}{k_{0}}-\alpha\right) + O(g_{{\rm eff}}^{4})\right]\\
&\Delta_{T}^{2}(k) =\Delta_{S}^{2}(k_{0}) \ r \left[1 + T\frac{k_{0}}{k} e^{\alpha}\left(\ln\frac{k}{k_{0}} - \alpha\right) + O(g_{{\rm eff}}^{4})\right] \nonumber
\end{align}
where $S$ and $T$ are constants,  $S=-D_{B}g_{{\rm YM}}^{2}(k^{*})N/k^{*}$ and $T=-D_{A}g_{{\rm YM}}^{2}(k^{*})N/k^{*}$.  The scale $k^{*}$ has been rewritten as $k^{*} = k_{0} e^{\alpha}$, where $k_{0}$ is the pivot scale at which the power spectra are compared with the data. In this form, $\alpha$ represents a shift in the coordinate $\ln(k/k_{0})$. The pre-factors $\Delta_{S}^{2}(k_{0})$ and $\Delta_{T}^{2}(k_{0}) = \Delta_{S}^{2}(k_{0}) \ r$ are the amplitudes of scalar and tensor perturbations, respectively, evaluated at the pivot scale.

A first consistency check can be made at this point, regarding the assumption of the large $N$ limit. In fact, from the COBE normalisation, $\Delta_{S}^{2}(k_{0}) \sim 10^{-9}$, leading to $N\sqrt{ C_{B}} \sim 2500$, which can only be justified by a large value of $N$, as was assumed.

As mentioned, these functions are both highly variable with $k$, mainly due to the $1/k$ behaviour in expression (\ref{lalala}), which comes from the running of $g_{{\rm eff}}$. This means that for sufficiently small values of $k$, scale invariance will break and they will present large tilts and runnings of the tilts. The deviation from scale invariance is highly constrained by observations so comparison with data should impose strong limits on the power spectra.

For the simplest case where the scale $k^{*}$ matches the pivot scale, \textit{i.e.} if $\alpha=0$, for a given pivot scale, the constant $S$ completely defines the shape of the scalar power spectrum and therefore the deviation from scale invariance at this pivot scale.  In fact, in this case $S=n_{s0}-1 + O(g_{{\rm eff}}^{4}(k_{0}))$, the spectral index at the pivot scale.

Of course that this matching between scales has no physical meaning. The pivot scale $k_{0}$ is just a scale used for practical reasons for data fitting whereas $k^{*}$ is the scale at which IR effects become non negligible, \textit{i.e.}, a parameter of the power spectra, and so a free parameter in CosmoMC. In this case, for simplicity, $\alpha$ is the free parameter. 

%

Since $\alpha$ corresponds to a shift in the coordinate $\log(k/k_{0})$ , allowing for $\alpha \neq 0$ can be viewed as allowing for a shift of the full power spectrum with respect to the pivot scale.  If $\alpha > 0$ one can see the power spectrum, and its features of scale invariance breaking, stretching towards larger values of $k$. If $\alpha < 0$, these features are squished towards smaller values of $k$. The spectral index at the pivot scale has the more general form:
\begin{align}
(n_{s0} - 1) &= \left. \frac{d \ln \Delta_{S}^{2}(k)}{d \ln k} \right|_{k=k_{0}} \\ 
&= \frac{S e^{\alpha} (1+\alpha)}{1-Se^{\alpha}\alpha} +  O(g_{{\rm eff}}^{4}(k_{0}))
\end{align}

Another consistency check needs to be made, regarding the choice of perturbative analysis in the dual theory. Since $|D_{B}| \sim 1$,
\begin{align}
g_{{\rm eff}}^{2}(k_{0})  \simeq  \abs{\frac{n_{s0}-1}{1+\alpha + \alpha(n_{s0}-1)}}
\label{gym}
\end{align} 
It is then necessary to check that this expression is always small for all values of $k$ and $\alpha$ probed by observations. This will be analyzed shortly.

The running of the spectral index can be expressed as a combination of the spectral index itself. 
\begin{align}
&{\rm running} = \left. \frac{d (n_{s} - 1)}{d \ln k} \right|_{k=k_{0}}\\ \nonumber
&= -2 (n_{s0} -1) - (n_{s0} - 1)^{2} + \frac{S e^{\alpha} \alpha}{1+S e^{\alpha}}  +O(g_{{\rm eff}}^{4}(k_{0}))
\end{align}\\

It is important to mention that if $S$ and $T$ are positive, the power spectra become negative for sufficiently small $k$. This is related to the error in $O(g_{\rm{eff}}^{2} \sim k^{-2})$. In our case, as will be seen, this presents no problem since the regime where the power spectra becomes negative naturally corresponds to scales much larger than can be studied by the CMB. 

The tensor to scalar ratio, $r$, can be used to obtain a constraint on the field content.
In this case, 
\begin{align}
r=\frac{\Delta_{T}^{2}(k_{0})}{\Delta_{S}^{2}(k_{0})} = \frac{\mathcal{N}_{A}+\mathcal{N}_{\phi}}{\mathcal{N}_{A}+\mathcal{N}_{\phi}+\mathcal{N}_{\chi}+2\mathcal{N}_{\psi}}
\end{align}

We made the comparison with data for two different pivot scales, $k=0.05  {\rm Mpc}^{-1}$ and $k=0.002   {\rm Mpc}^{-1}$ to probe different regions of the power spectra.

\begin{figure}[h]
\vspace*{-0.5cm}
\includegraphics[width=0.66\linewidth]{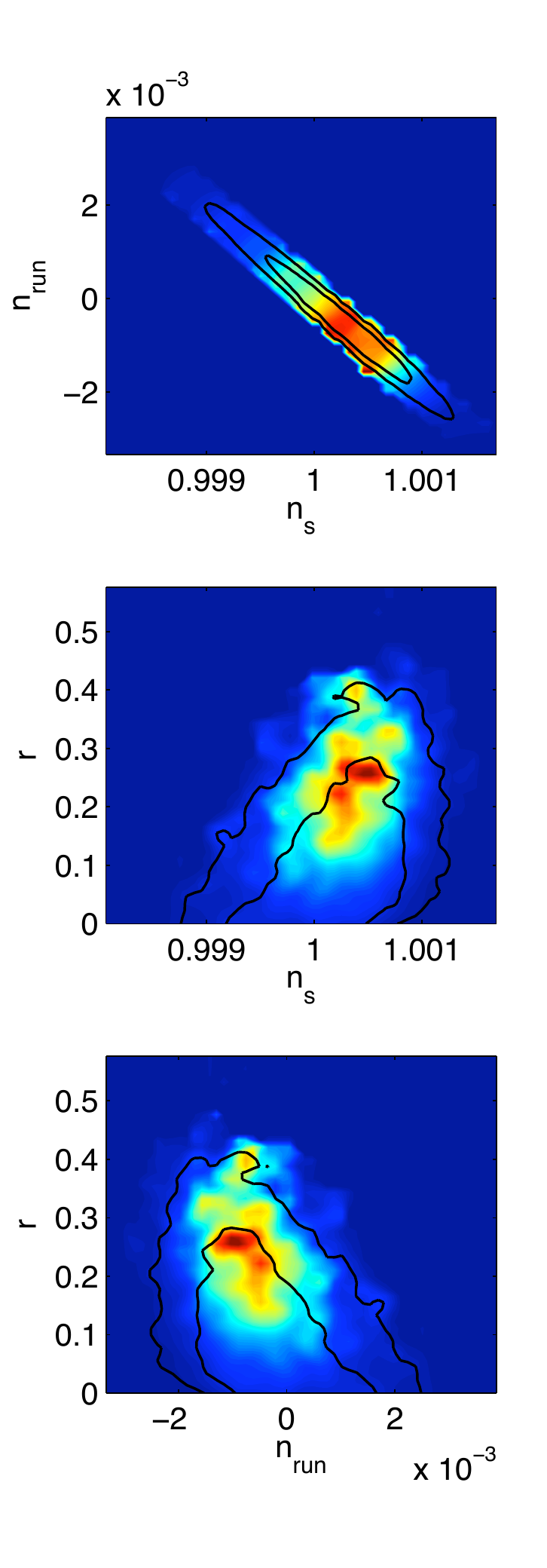}
\vspace*{-1cm}
\caption{Best-fit parameters for the pivot scale $k_{0}=0.05 \rm{Mpc}^{-1}$, when $\alpha=0$. The contours show the  $68\%$ and $95\%$ confidence limits, for $n_{s0}$, the running $n_{{\rm run}}$ and tensor to scalar ratio $r$. The fit was done for WMAP 7 data alone \cite{wmap}. }
\label{noa005}
\end{figure}

\begin{figure}
\vspace*{-0.5cm}
\includegraphics[width=0.7\linewidth]{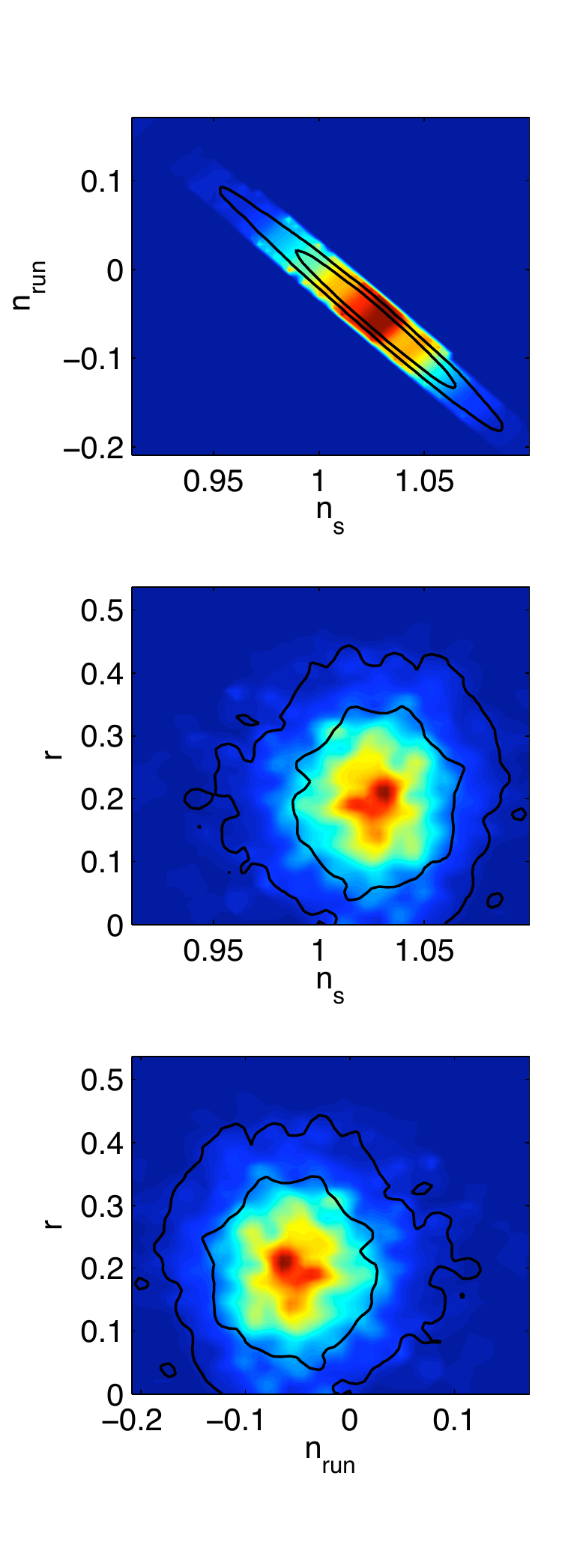}
\vspace*{-1cm}
\caption{Best-fit parameters for the pivot scale $k_{0}=0.002  {\rm Mpc}^{-1}$, when $\alpha=0$. As in figure \ref{noa005}, the contours show the  $68\%$ and $95\%$ confidence limits, for $n_{s0}$, the running $n_{{\rm run}}$ and tensor to scalar ratio $r$.}
\label{noa0002}
\end{figure}

\begin{figure}
\includegraphics[width=1\linewidth]{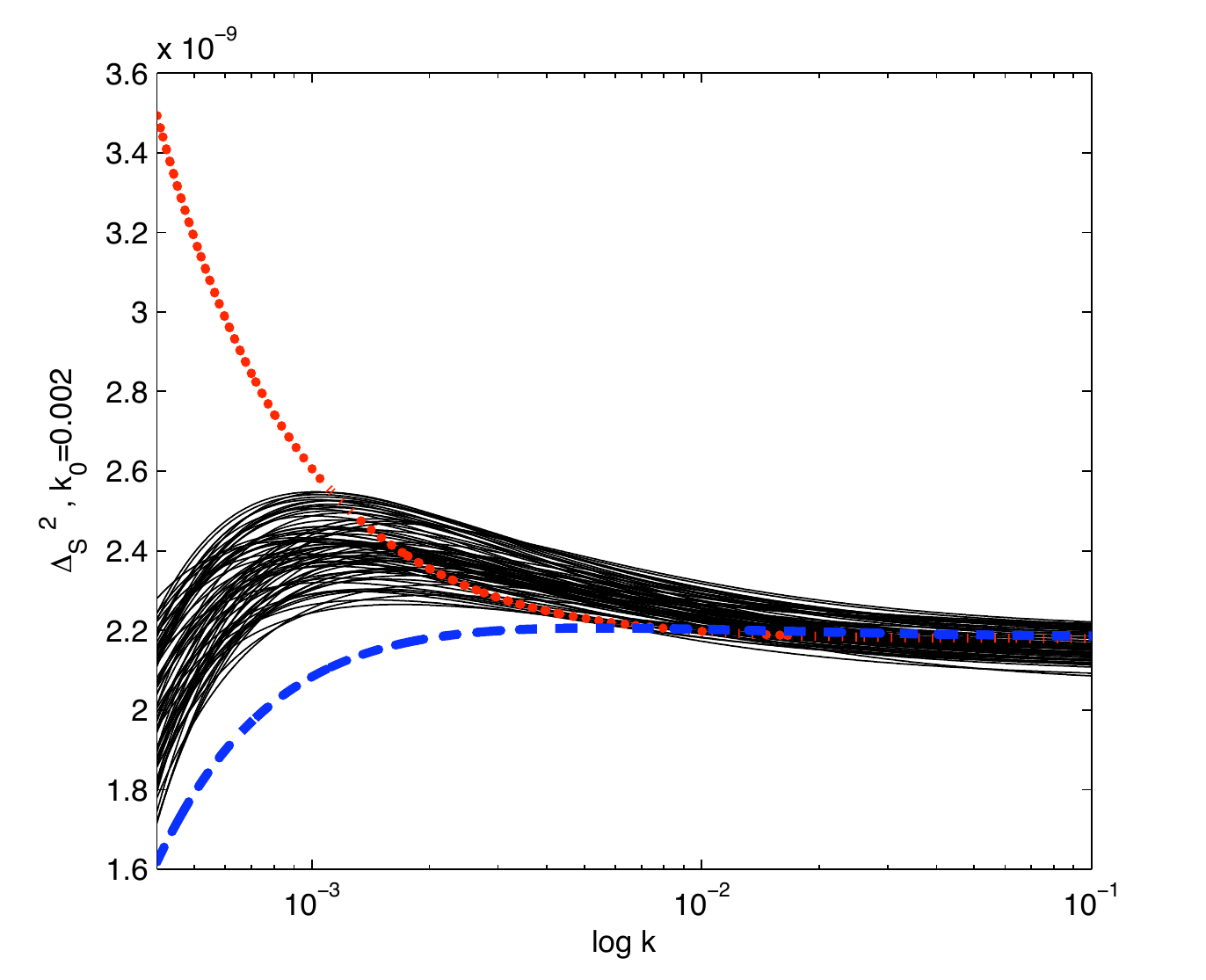}
\caption{Power spectrum of scalar perturbations. The dashed (blue) line represents the best fit curve for the model with $\alpha=0$, with $k_{0}=0.002  {\rm Mpc}^{-1}$. The dotted (red) line shows how the power spectrum with $\alpha=0$ would look like if it presented $(ns_{0}-1)=-0.001$ at $k_{0}=0.005 {\rm Mpc}^{-1}$; the features at low $k$ are in disagreement with data. The black lines represent the best fit curves for the model with $\alpha \neq 0$ for $k_{0}=0.002 {\rm Mpc}^{-1} $.}
\label{ps}
\end{figure}

For $k=0.05  {\rm Mpc}^{-1}$ and $\alpha=0$ we find that the data forces the power spectrum of density perturbations to be extremely flat, figure \ref{noa005}. 
For $k=0.002  {\rm Mpc}^{-1}$ and $\alpha=0$ we can see that this pivot scale is already near the regime where scale invariance breaks, with a preference for positive $(n_{s0} -1)$, figure \ref{noa0002}. 
It is known, from the standard WMAP analysis using a Taylor expansion type of density power spectrum, that the data has a slight preference for $n_{s0} < 1$ at $k=0.05  {\rm Mpc}^{-1}$ \cite{wmap}. With the present power spectrum, because of the large running, a small deviance from scale invariance at this scale, would mean a strong disagreement with data at lower values of $k$. In figure \ref{ps} this effect is shown. The dashed and dotted (blue and red) lines correspond to the case of $\alpha=0$. The dashed line (blue) shows $\Delta_{S}^{2}(k)$ for the best fit parameters at $k_{0}=0.002  {\rm Mpc}^{-1}$; it is easy to see how flat it needs to be to agree with data.  The dotted line (red) shows how $\Delta_{S}^{2}(k)$ would look if $n_{s0}-1 = -0.001$ at $k_{0}=0.05  {\rm Mpc}^{-1}$; it is possible to see how the strong running destroys the fitting with data.\\

\begin{figure}[h]
\vspace*{-1cm}
\includegraphics[width=0.7\linewidth]{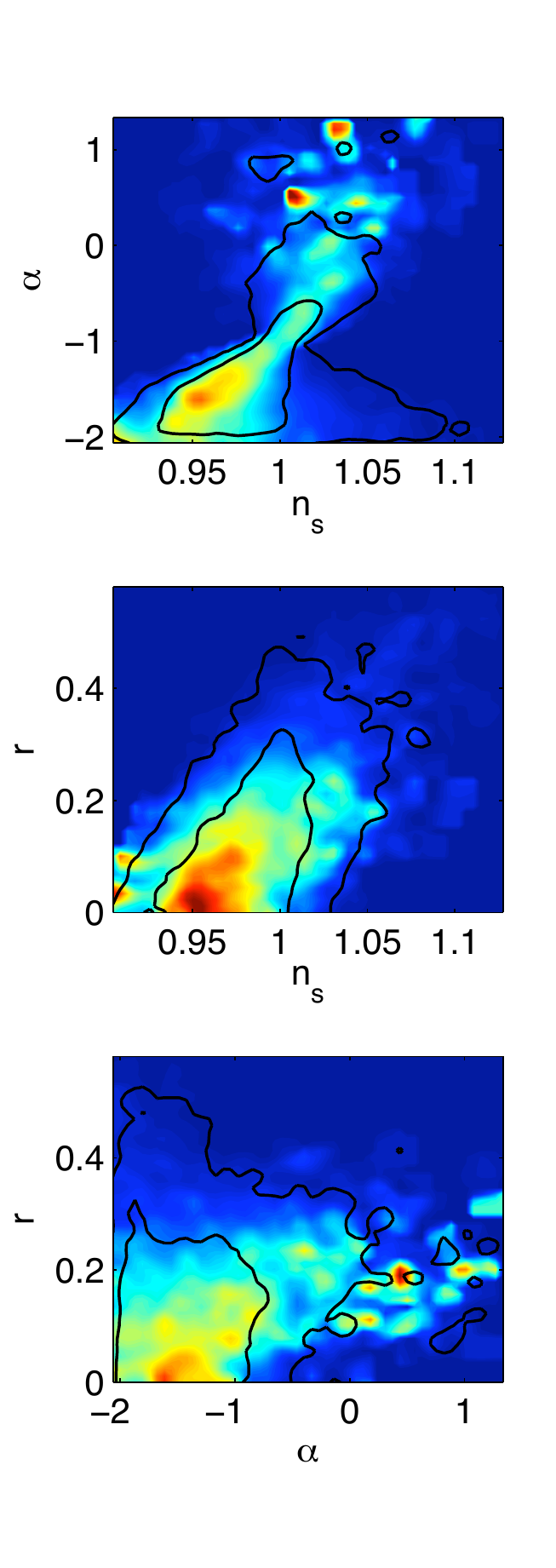}
\vspace*{-1cm}
\caption{Best-fit parameters for the pivot scale $k_{0}=0.002  {\rm Mpc}^{-1}$ with $\alpha$ in the range [-2, 2]. Contours and data as in figures \ref{noa005} and \ref{noa0002}.}
\label{witha0002v1}
\end{figure}

\begin{figure}
\vspace*{-1cm}
\includegraphics[width=0.68\linewidth]{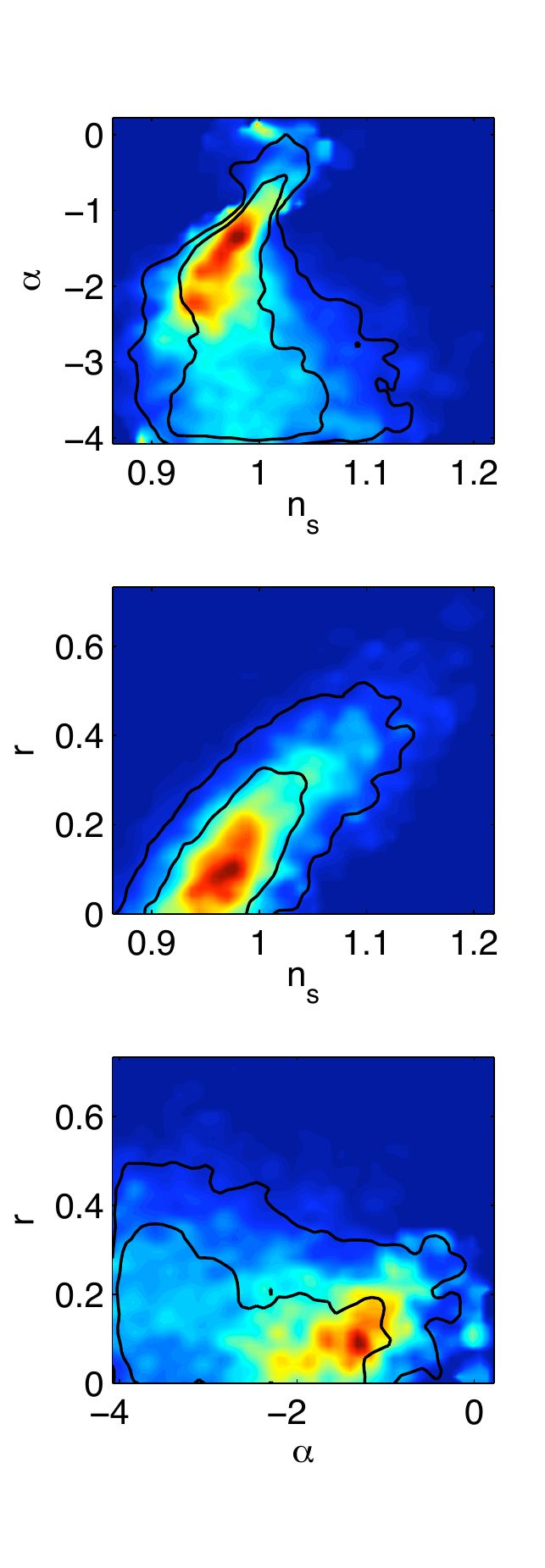}
\vspace*{-1cm}
\caption{Best-fit parameters for the pivot scale $k_{0}=0.002  {\rm Mpc}^{-1}$ with $\alpha$ in the range [-4, 4]. Contours and data as in figures \ref{noa005} and \ref{noa0002}.}
\label{witha0002v2}
\end{figure}

When $\alpha$ is allowed to be non-zero, data forces it to always have negative values, as can be seen in figures \ref{witha0002v1} and \ref{witha0002v2}. As mentioned, this has the effect of shifting the break of scale invariance features towards lower values of $k$. The best-fit is obtained for $\alpha \sim -1.4$. In this case, the effect of $\alpha$ is such that, for the appropriate value of $S$, the power spectrum can be chosen to fit some small negative tilt preferred by data at almost all scales, as well as the low value for the quadrupole. This is shown by the black lines in figure \ref{ps}; in this figure only the best likelihood curves are plotted. At $k=0.05  {\rm Mpc}^{-1}$ the power spectrum is still extremely near scale invariance. 

\begin{figure}
\vspace*{-1cm}
\includegraphics[width=0.7\linewidth]{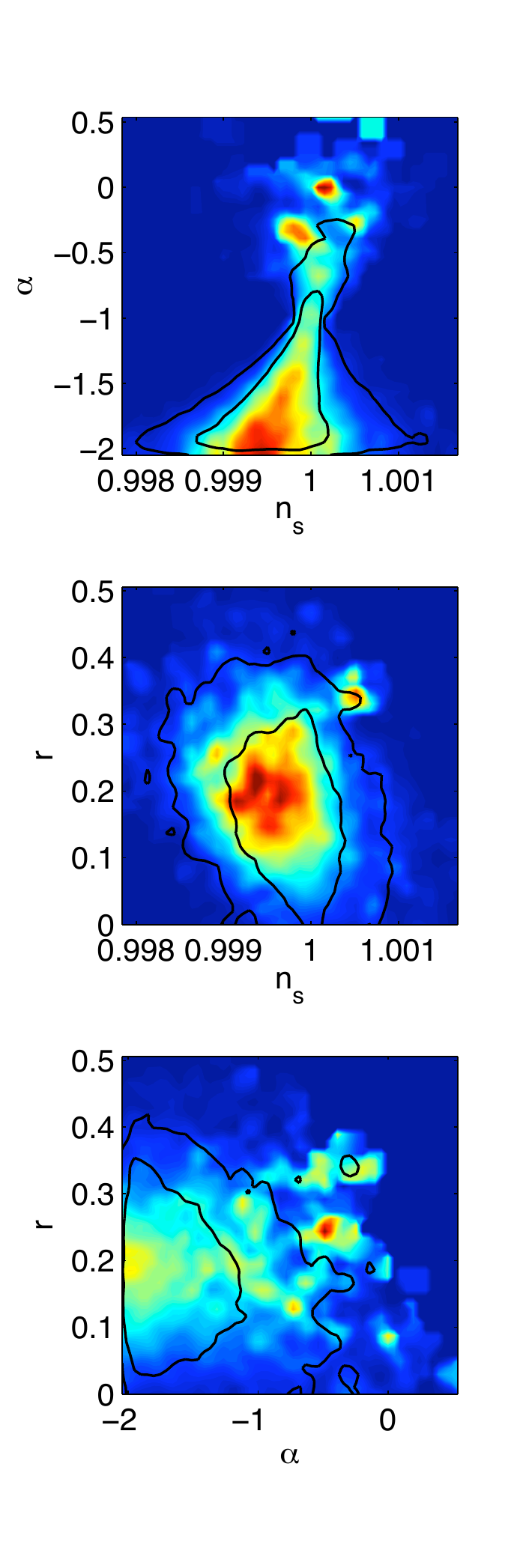}
\vspace*{-1cm}
\caption{Best-fit parameters for the pivot scale $k_{0}=0.05 {\rm Mpc}^{-1}$ with $\alpha$ in the range [-2, 2]. Contours and data as in figures \ref{noa005} and \ref{noa0002}.}
\label{witha005}
\end{figure}

The best-fit remains the same when $\alpha$ is allowed to take a wider range of negative values, as can be seen in figure \ref{witha0002v2}. However, all possible negative values of $\alpha$ are allowed by the data. In the case of very negative $\alpha$, the features in the power spectrum are pushed towards values of $k$ smaller than the quadrupole, and so unconstrained by data, relaxing the allowed values of $n_{s0}$, even if keeping it extremely small for small scales, as can be seen for the case of $k_{0}=0.05  {\rm Mpc}^{-1}$ in figure \ref{witha005}. These solutions for the power spectrum fit the data even if with lower likelihood than the best-fit. 

In terms of the meaning of $k^{*}$, these results just say that data is in agreement with any value for the scale at which IR corrections become important in the perturbative analysis, provided that this is equal or smaller than the wavenumbers probed by the CMB. \\

At this stage it is possible to try to constrain the field content of the dual boundary theory. As seen, the field content is directly related to the tensor to scalar ratio. Since, as can be seen in figures \ref{witha0002v1} and \ref{witha0002v2}, $r  \lesssim 0.5$ at 95$\%$ confidence level,
\begin{align}
\mathcal{N}_{A}+\mathcal{N}_{\phi} \lesssim \mathcal{N}_{\chi}+2\mathcal{N}_{\psi}
\end{align}
which is far from being a strong constraint on the field content. However, future experiments such as Planck, may impose considerably stronger limits on the value of $r$. In that case, since the allowed field content is very sensitive to $r$, this relation might become much more interesting. For example, a stronger upper limit, would mean that a nearly symmetric field content is excluded by data.\\

\begin{figure}[h]
\includegraphics[width=0.7\linewidth]{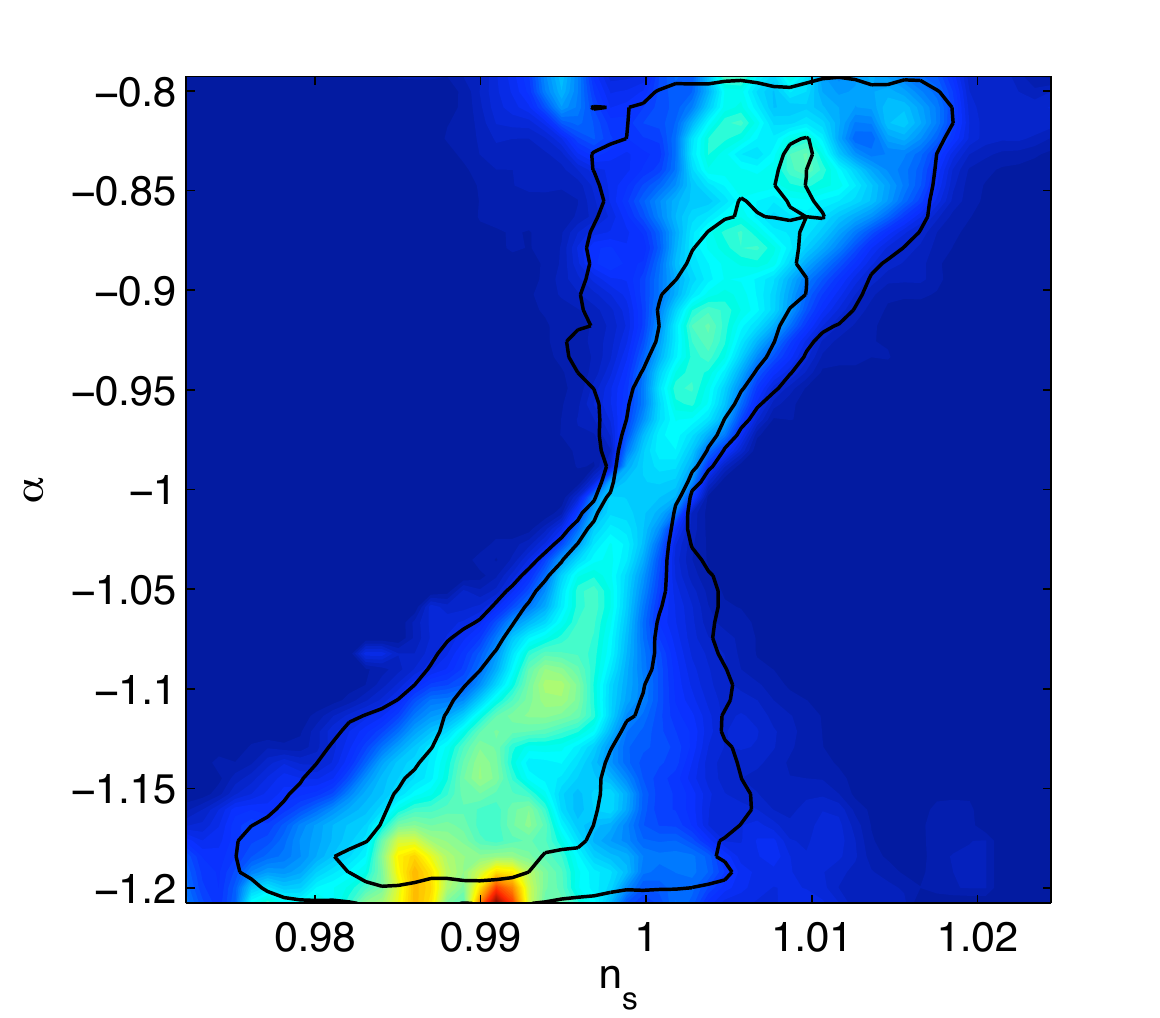}
\caption{Best-fit parameters for the pivot scale $k_{0}=0.002 {\rm Mpc}^{-1}$ with $\alpha$ in the range [-1.2, -0.98]. Contours and data as in figures \ref{noa005} and \ref{noa0002}.}
\label{witha0002v3}
\end{figure}

It is now necessary to check that the value of  $g_{{\rm eff}}^{2}(k_{0})$ is always small for the scales of interest of the CMB, as was assumed by making the perturbative expansion on the boundary field theory. 

For small scales, or large values of $k$, the power spectrum is always very flat and so $|n_{s0}-1|$ is very small. This is the case for $k$ larger than $k=0.002 {\rm Mpc}^{-1}$; at this particular scale the value of $|n_{s0}-1|$ is always small and takes a maximum value of $\sim 0.1$. For these scales, expression (\ref{gym}) can only become critically large around $\alpha \sim -1$. As can be seen in figure \ref{witha0002v1} and figure \ref{witha0002v2}, data forces $n_{s0}-1$ to be extremely close to zero around $\alpha \sim -1$; the question is if, in fact, $n_{s0}-1$ decreases faster around this point than $1+\alpha+(n_{s0}-1)\alpha$. 

As can be seen in figure \ref{witha0002v3}, interestingly, the maximum value that $g_{\rm eff}^{2}(k=0.002)$ can take at each $\alpha$ is independent of $\alpha$. The slope of the $95 \%$ confidence level around $\alpha = -1$ is approximately constant. For this scale, $g_{\rm eff}^{2}$ can take a maximum value of 0.1, which is barely in agreement with the perturbative treatment at the boundary. In this figure, it isn't clear that $n_{s0}$ actually converges to 1 for $\alpha=-1$ but that actually happens; any broadening of the contours is related to the smoothing of the plotting. 

This behaviour around $\alpha=-1$ occurs for every scale, and for $k > 0.002 {\rm Mpc}^{-1}$ the maximum value that $g_{\rm eff}^{2}$ is allowed to take is even smaller and so more comfortably in agreement with the perturbative calculations. \\

For smaller values of $k$ problems arise. The CMB data probes scales up to $\sim 10^{-3} {\rm Mpc}^{-1}$ and in fact, for models for which $g_{\rm eff}^{2}(k=0.002  {\rm Mpc}^{-1}) \sim 0.1$, $g_{\rm eff}^{2}(k=0.0002 {\rm Mpc}^{-1}) \sim 0.1 \times 10 = 1$.
This is in disagreement with the loop perturbative treatment since, in this case, the 3-loop and higher corrections would be of the same order of magnitude as the 2-loop one. Clearly, at these scales, the calculations at 2-loop order are insufficient. Then, there is a limit for the validity of the studied power spectra, that can be identified to be at $\sim k=0.002 {\rm Mpc}^{-1}$; for smaller $k$, higher corrections should be taken into account. 

Data from very large scales, like the quadrupole, can not be used to probe this model, as constructed in this work. So, for the range of scales where the model is valid, it can be concluded the data fits the parameters to an almost scale invariant density power spectrum, with a maximum tilt of $|n_{s0}-1| \sim 0.1$ at $k_{0{\rm min}}=0.002  {\rm Mpc}^{-1}$.

\section{Conclusions}

The aim of this work was to constrain the scenario of a strongly coupled gravity at early times by studying the cosmological inflation in the dS/QFT framework and fitting it with data.

The dS/QFT correspondence relates correlation functions of a bulk inflation theory with some dual correlation functions of a field theory at the boundary of the de Sitter inflationary space. This duality is such that if the bulk theory is strongly coupled, the boundary one is weakly coupled and vice versa. The idea developed in this paper is to study the dynamics of inflation in its dual boundary theory, assuming a perturbative treatment to be valid. In this case, the bulk gravity theory is strongly coupled and cannot be described by geometrical fluctuations of the background metric. The observed power spectra of metric perturbations can then be expressed in terms of the dynamics of the boundary field theory.

Since the field theory dual to 4-$D$ de Sitter spaces is unknown, an \textit{ad-hoc} Yang-Mills theory with a large freedom of parameters was introduced. The strategy was, by computing and comparing the power spectra with WMAP 7 year release data, to constrain the space of possible parameters and in this way constrain the whole scenario of strongly coupled gravity at early times.

The correlation functions used were corrected up to the 2-loop order. The fit with CMB data allowed us to identify the regime of validity of such a truncation. It turns out that the resulting power spectra are only valid to describe perturbations for the CMB scales with $k \gtrsim 0.002 {\rm Mpc}^{-1}$. 

For this range of scales, it can be concluded that the model is barely constrainable at present, but that it is nevertheless quite predictive.

In terms of observable parameters, the results showed that for large $k$ the power spectra need to be very flat. In particular, the spectral index of the density power spectrum for the pivot scale $k_{0}=0.005  {\rm Mpc}^{-1}$ needs to be within $|n_{s0}-1| \lesssim0.002$, with a preference for negative $n_{s0}$. As $k$ decreases, the interval of allowed values that $n_{s0}$ can take is relaxed and, for example for $k_{0}=0.002  {\rm Mpc}^{-1}$, $|n_{s0}-1| \lesssim 0.1$. 

The tensor to scalar ratio is constrained to be $\lesssim 0.5$. 

In terms of the free parameters of the model, these results represent almost no information. The field content needs to respect the not very restrictive relation
\begin{align}
\mathcal{N}_{A}+\mathcal{N}_{\phi} \lesssim \mathcal{N}_{\chi}+2\mathcal{N}_{\psi}
\end{align}
and the scale $k^{*}$ can be any, provided that it is of the same order or smaller than the wavenumbers probed by observations of the CMB. However, much more interesting constraints on the field content are expected from the comparison with data from future experiments like Planck, for which stronger limits on the value of $r$ are forecasted. \\

This model, however, presents a strong conceptual problem regarding the transition from the strongly coupled gravity theory at early times to the weakly coupled hot Big Bang. In other words, there is the lack of a satisfactory mechanism to end inflation. The connection between these two different phases of evolution remains an important open question.


\begin{acknowledgments}
I am particularly grateful to Andrew Liddle and David Seery for fundamental help in this work. I would like to thank Paul McFadden and Kostas Skenderis for useful comments. This work was supported by FCT (Portugal). 
\end{acknowledgments}


\end{document}